\documentclass[twocolumn,showpacs,showkeys,prl]{revtex4}
\usepackage{graphicx}
\usepackage{dcolumn} 
\newcommand{\st}{SrTiO$_3$}
\newcommand{\bt}{BaTiO$_3$}

\begin{document}

\title{Spin singlet small bipolarons in Nb-doped  \bt}
\author{T. Kolodiazhnyi}
 \email{koloditv@mcmaster.ca}
\author{S. C. Wimbush}
\affiliation{National Institute for Materials Science, ICYS, 1-1
Namiki, Tsukuba, Ibaraki 305-0044, Japan}

\date{\today}

\begin{abstract}

The magnetic susceptibility and electrical resistivity of n-type
BaTi$_{1-x}$Nb$_x$O$_3$ have been measured over a wide temperature
range. It is found that, for 0 $<$ \emph{x} $<$ 0.2, dopant
electrons form immobile spin singlet small bipolarons with binding
energy around 110 meV. For \emph{x} = 0.2, a maximum in the
electrical resistivity around 15 K indicates a crossover from band
to hopping transport of the charge carriers, a phenomenon expected
but rarely observed in real polaronic systems.

\end{abstract}

\pacs{71.38.Mx, 71.38.Ht}
\keywords{\bt, bipolarons, magnetic properties, electrical
resistivity}
\maketitle


The theoretical treatment of itinerant carriers interacting with a
strongly polarizable lattice has been conducted for over 50 years,
resulting in a variety of different models. Depending on the degree
of electron-lattice interaction, these models can be roughly divided
into two main groups. The first \cite{pekar:1946,frohlich:1954}
treats the electron as a Bloch-wave-like delocalized particle called
a Fr\"{o}hlich or large polaron. The second
\cite{tjablikov:1952,holstein:1959} assumes an extreme localization
of the itinerant electron within one or several lattice sites (i.e.,
a small polaron, SP) accompanied by a strong local lattice
deformation. It is expected that under certain conditions, polarons
can pair to form large/small bipolarons, BP
\cite{vinetskii:1957,emin:1989}. Although not without controversy
\cite{chakraverty:1998}, several authors suggest \cite{emin:1989}
that Bose-Einstein condensation of bipolarons may result in
high-\emph{T}$_c$ superconductivity.

Evidence of a strong electron-lattice interaction has been
accumulating for \emph{A}TiO$_3$, where \emph{A} = Ba or Sr
\cite{levstik:2002,lenjer:2002}. These are textbook examples of
crystals having a simple perovskite structure and high static
dielectric constant furnished by a soft phonon mode. Stoichiometric
\emph{A}TiO$_3$ are band gap insulators. Donor doping or reduction
in an oxygen-deficient atmosphere results an n-type material with
unconventional electronic conductivity where the carriers move
within a very soft and deformable lattice. In \st, the overlap
between the Ti-Ti 3d t$_{2g}$ orbitals is strong enough to prevent
the localization of the itinerant electrons which then form large
polarons. As a result, n-type \st\ shows metallic conductivity with
an enhanced effective electron mass \emph{m$^*$} of 6\emph{m$_e$} to
14\emph{m$_e$} \cite{gervais:1993,verbist:1992}. It has been argued
that, at low temperatures, large polarons in n-type \st\ condense
into the large BPs responsible for superconductivity below 0.6 K
\cite{verbist:1992}.

The lattice constant of cubic \bt\ (\emph{a} = 0.4005 nm) is 2.5$\%$
larger than that of \st\ (\emph{a} = 0.3905 nm). This small increase
in \emph{a} causes a dramatic difference in the electronic
properties of the two compounds. Unlike n-type \st, the conductivity
of n-type \bt\ shows insulating behavior with the low conductivity
attributed to the thermally activated hopping of strongly localized
electrons (i.e., small polarons) between neighboring Ti sites.
Gerthsen \emph{et al}.\cite{gerthsen:1965} explained the
mid-infrared reflectance of n-type \bt\ in terms of the SP
absorption model. In accord with the hopping transport of SPs,
Bursian \emph{et al}.\cite{bursian:1972} found that the Seebeck
coefficient of weakly-doped n-type \bt\ is temperature independent
above 300 K. Direct evidence of the self-trapped Ti$^{3+}$ SP was
provided by ESR studies of acceptor depleted \bt\
\cite{scharfschwerdt:1996}. Nevertheless, the SP interpretation of
the electronic properties of \bt\ is far from uncontroversial. Two
major points of disagreement are the similarity in the Hall and
drift mobilities ($\mu_{\mathrm{Hall}}$ $\approx$
$\mu_{\mathrm{drift}}$ $\approx$ 0.2 cm$^2$/V s at 300 K) which,
according to the SP model, should satisfy $\mu_{\mathrm{Hall}}$
$\gg$ $\mu_{\mathrm{drift}}$ \cite{ihrig:1978}, and the large
difference between the optical and thermal activation energies
(\emph{E}$_{\mathrm{op}}$ = 500 -- 600 meV, \emph{E}$_{\mathrm{th}}$
= 23 meV) of the proposed SPs, that should satisfy
\emph{E}$_{\mathrm{op}}$ = 4\emph{E}$_{\mathrm{th}}$
\cite{ihrig:1976}.

In our opinion, the polaronic approach to understanding the unusual
properties of n-type \bt\ requires that the possibility of electron
pairing be taken into account. The coupling of electrons into
Anderson-type \emph{on-site} spin singlet Ti$^{2+}$ small bipolarons
according to Ti$^{4+}$ + 2\emph{e}$^-$ $\rightarrow$ Ti$^{2+}$ was
first postulated by Moizhes and Suprun \cite{moizhes:1984}. Later,
the idea of BPs in \bt\ was supported by Lenjer and co-workers
\cite{lenjer:2002}, who observed an unexpected increase in the
Ti$^{3+}$ ESR signal with temperature. This observation led them to
suggest that, due to the high correlation energy that sets the
energy cost for a double occupancy of the Ti$^{2+}$ ion, the
symmetry of the BP is rather of the \emph{inter-site}, i.e.,
Heitler-London type, where the electron pair is localized at two
neighboring Ti$^{3+}$ ions according to 2Ti$^{4+}$ + 2\emph{e}$^-$
$\rightarrow$ 2Ti$^{3+}$.

The central question addressed in this Letter is whether the
Ti$^{3+}$ \emph{S} = $\frac{1}{2}$ small polarons in n-type \bt\
remain isolated at \emph{T} = 0 K or if they pair into spin singlet
\emph{S} = 0 bipolarons. We also report on the physical properties
of the BPs in n-type \bt, as revealed by magnetic and electrical
measurements.

Polycrystalline BaTi$_{1-x}$Nb$_x$O$_3$ samples with 0 $\le$
\emph{x} $\le$ 0.2 were prepared by solid state reaction from high
purity 99.99+$\%$ BaCO$_3$, TiO$_2$, and Nb$_2$O$_5$. Sintering was
performed in forming gas (3$\%$ H$_2$/97$\%$ N$_2$) to ensure
electronic rather than cation vacancy compensation of the Nb dopant
\cite{lewis:1986}. Phase purity of the ceramics was confirmed by
powder x-ray diffraction and energy dispersive x-ray microanalysis
with no evidence of secondary phases up to \emph{x} = 0.2. According
to x-ray photoelectron spectroscopy measurements, charge
compensation in the sintered BaTi$_{1-x}$Nb$_x$O$_3$ samples is
accomplished by the formation of Ti$^{3+}$ ions while the Nb ions
remain predominantly in the 5+ oxidation state.

\begin{figure}[htb]
 \begin{center}
\scalebox{1.1}{\includegraphics [0,0][176,189]{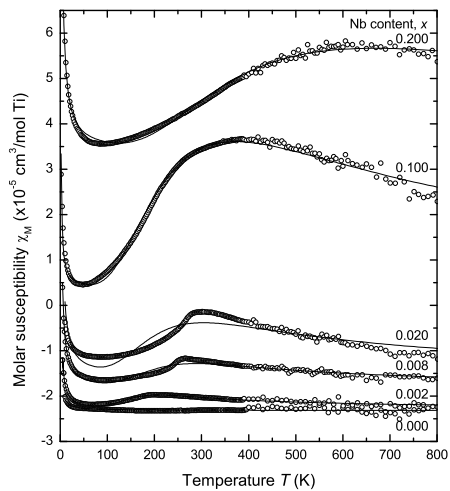}}
  \caption{Magnetic susceptibility per mole of Ti of the BaTi$_{1-x}$Nb$_x$O$_3$ ceramics with \emph{x} = 0, 0.002, 0.008, 0.020,
  0.100 and 0.200.
  Solid lines are fits to the data as described in the text.}
  \label{fig1}
 \end{center}
\end{figure}

Assuming strong localization of excess electrons on isolated
Ti$^{3+}$ ions down to 0 K (SP ground state), the magnetic
susceptibility should decrease with temperature in accordance with
the Curie law,
\begin{equation}
\chi_{\mathrm{para}} = \frac{N_{\mathrm{A}} p^2 \mu_{\mathrm{B}}^2
c_{\mathrm{P}}}{3k_{\mathrm{B}}T}, \label{eq1}
\end {equation}
where \emph{N}$_{\mathrm{A}}$ is the Avogadro number, $p$ is the
effective magnetic moment in Bohr magnetons ($\mu_{\mathrm{B}}$),
\emph{k}$_{\mathrm{B}}$ is the Boltzmann constant, \emph{T} is the
temperature and \emph{c}$_{\mathrm{P}}$ is the concentration of
small polarons. Figure \ref{fig1} shows the temperature dependence
of the molar magnetic susceptibility, $\chi$$_{\mathrm{M}}$, of the
BaTi$_{1-x}$Nb$_x$O$_3$ ceramics in the 2 -- 800 K temperature
range. It is obvious that the $\chi$$_{\mathrm{M}}$(\emph{T}) data
do not support a simple SP scenario.
\begin{figure}[htb]
 \begin{center}
\scalebox{1.1}{\includegraphics*{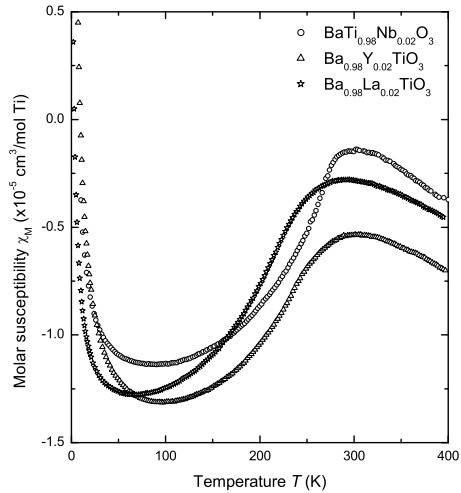}}
  \caption{Comparison of the magnetic susceptibility of Nb-doped \bt\ with that of \bt\ doped with donors having a stable oxidation state (Y$^{3+}$ and La$^{3+}$).
  All three curves reveal a magnetic anomaly.}
  \label{fig1b}
 \end{center}
\end{figure}
The observed anomalous increase in $\chi$$_{\mathrm{M}}$ at
intermediate temperatures is associated with the breaking up of spin
singlet BPs. An overall increase in $\chi$$_{\mathrm{M}}$ with
dopant concentration is attributed to the formation of the BP band
with associated van Vleck paramagnetism. To fit the data we apply
Emin's formalism \cite{emin:1996} for dissociation of small BPs in a
weak magnetic field (\emph{g}$\mu_{\mathrm{B}}$\emph{B} $\ll$
2$k_{\mathrm{B}}T$). According to Emin, the concentration of SPs,
\emph{c}$_{\mathrm{P}}$, produced by thermal dissociation of BPs is
given by
\begin{equation}
c_{\mathrm{P}} = \frac{1-(1-x)\sqrt{1+ \left[ x(2-x)/(1-x)^2 \right]
e^{\varepsilon_{\mathrm{b}}/k_{\mathrm{B}}T}
}}{1-e^{\varepsilon_{\mathrm{b}}/k_{\mathrm{B}}T}}, \label{eq2}
\end {equation}
where \emph{x} is the concentration of the Nb$^{5+}$ dopant and
$\varepsilon$$_{\mathrm{b}}$ is the BP binding energy, i.e., the
energy required to split the BP into two SPs. The data in
Fig.\ref{fig1} were fitted with the general formula
\begin{equation}
\chi_{\mathrm{M}} = \chi_{\mathrm{dia}} + \chi_{\mathrm{VV}} +
\frac{A}{T} + \chi_{\mathrm{para}}, \label{eq3}
\end {equation}
where $\chi_{\mathrm{dia}}$ = --2.32$\times$10$^{-5}$ cm$^3$/mol is
the diamagnetic susceptibility of the undoped, stoichiometric \bt,
$\chi_{\mathrm{VV}}$ is the temperature independent van Vleck
paramagnetic contribution,  and \emph{A} accounts for
impurity-related paramagnetism dominant at \emph{T} $<$ 40 K.

The van Vleck term is given by
\begin{equation}
\chi_{\mathrm{VV}}=\frac{2N_{\mathrm{A}} x |\langle l
|\mu_{\mathrm{z}}|0\rangle|^2}{\Delta}
 \label{eq4}
\end {equation}
where $\langle l |\mu_{\mathrm{z}}|0\rangle$ is a non-diagonal
matrix element of the magnetic moment operator connecting the BP
ground state 0 with the excited state \emph{l} of energy $\Delta$ =
\emph{E$_l$} -- \emph{E$_0$} above the ground state.

The results of the fit are shown as solid lines in Fig.\ref{fig1}.
The fitting parameters are summarized in Table \ref{table1}. A
somewhat larger value of \emph{A} for the \emph{x} = 0.2 sample is
attributed to the partial compensation of the Nb dopant by
paramagnetic cation vacancies as we approach the solubility limit of
Nb in \bt. In agreement with Eq.\ref{eq4}, $\chi_{\mathrm{VV}}$ is
found to scale linearly with dopant concentration. Good fits were
obtained for \emph{x} = 0, 0.002, 0.1 and 0.2 and somewhat worse
fits for \emph{x} = 0.008 and 0.02. We can explain the poor fit at
intermediate dopant concentrations since, while fitting the data, we
have assumed a constant $\varepsilon$$_{\mathrm{b}}$ and
$\chi_{\mathrm{VV}}$ for the entire temperature range. It is known,
however, that undoped \bt\ exists in four crystallographic forms
\cite{jona:1962}: rhombohedral at \emph{T} $<$ 200 K, orthorhombic
at 200 K $<$ \emph{T} $<$ 270 K, tetragonal at 270 K $<$ \emph{T}
$<$ 393 K and cubic at 393 K $<$ \emph{T} $<$ 1670 K. Since both
$\varepsilon$$_{\mathrm{b}}$ and $\chi_{\mathrm{VV}}$ depend on the
crystal symmetry, the use of single values of
$\varepsilon$$_{\mathrm{b}}$ and $\chi_{\mathrm{VV}}$ for \emph{x} =
0.008 and 0.02 gives only marginally satisfactory results. At the
lowest dopant concentration (\emph{x} = 0.002), 70$\%$ of the BPs
are already dissociated into SPs at 200 K, i.e., fully within the
rhombohedral phase. As a result, this fit is good in spite of the
single values of $\varepsilon$$_{\mathrm{b}}$ and
$\chi_{\mathrm{VV}}$ used. Doping with Nb results in a significant
lowering of the phase transition temperatures in \bt\
\cite{feltz:1977}. Indeed, low-temperature x-ray analysis of the
\emph{x} $\ge$ 0.1 samples confirmed that they remain cubic down to
at least 20 K. Hence, the use of single values of
$\varepsilon$$_{\mathrm{b}}$ and $\chi_{\mathrm{VV}}$ produces a
good fit also in these cases. The data show that in the doping range
covering two orders of magnitude, the bipolaron binding energy is of
the order of 110 meV. For the most highly doped sample (\emph{x} =
0.2), the bipolaron binding energy (172 meV) is found to be rather
high. This may be due to the solubility limit of donor atoms and/or
BP overcrowding effects.
\begin{table}[htb]
\begin{center}
\caption {\label{table1}Fitting parameters for $\chi_{\mathrm{M}}$
of BaTi$_{1-x}$Nb$_x$O$_3$.}
\begin{tabular}{lcccc}\hline\hline\\[0.1ex]
\emph{x}  &$\chi_{\mathrm{VV}}$$\times$10$^5$  & \emph{A}$\times$10$^5$ & $\varepsilon$$_{\mathrm{b}}$ & \emph{p}\\
& [cm$^3$/mol]& [cm$^3$K/mol]& [meV]&  \\[1ex]
 \hline\\[0.5ex]
0.000 & 0    & 2.16 &  --  & --   \\[0.5ex]
0.002 & 0.02 & 5.73 & 106 & 0.57  \\[0.5ex]
0.008 & 0.47 & 11.73 & 122 & 0.46  \\[0.5ex]
0.020 & 0.73 & 16.64 & 111 & 0.46 \\[0.5ex]
0.100 & 2.68 & 5.96 & 105 & 0.42 \\[0.5ex]
0.200 & 5.69 & 22.5 & 172 & 0.36 \\[0.5ex]
 \hline
\end{tabular}
\end{center}
\end{table}

One might argue, however, that the BP interpretation of the
$\chi_{\mathrm{M}}$ anomaly is not convincing enough, and that
thermal activation of intrinsic or extrinsic lattice defects may
generate similar effects. Let us address this crucial argument in
detail. Since no $\chi_{\mathrm{M}}$ anomaly is observed in undoped,
stoichiometric \bt, the possibility of the \emph{S} = 0
$\rightarrow$ \emph{S} = $\frac{1}{2}$ activation of any supposed
extrinsic impurities can be ruled out. According to the point defect
chemistry of \bt, the Nb$^{5+}$ dopant is compensated by barium or
titanium vacancies at high oxygen partial pressures
(\emph{P}$_{\mathrm{O}_2}$) and by electrons at low
\emph{P}$_{\mathrm{O}_2}$ \cite{lewis:1986}. Since the samples were
prepared at low \emph{P}$_{\mathrm{O}_2}$, the concentration of
cation vacancies is negligible. However, even if a small amount of
cation vacancies were still present in the heavily doped samples,
their thermal activation energy is at least five times higher than
$\varepsilon$$_{\mathrm{b}}$ \cite{lewis:1986}. It is quite possible
that oxygen vacancies ($\mathrm{V_{O}}$) are present in the samples.
According to recent calculations \cite{honnerberg:2000}, these
vacancies will form a $[$Ti$^{3+}$--$\mathrm{V_{O}}$--Ti$^{3+}$$]$
neutral complex defect. In fact, the
$[$Ti$^{3+}$--$\mathrm{V_{O}}$--Ti$^{3+}$$]$ complex is nothing
other than the small BP bound to the oxygen vacancy, although in
this case the electrons are localized on the Ti 3z$^2$--r$^2$
orbitals \cite{solovyev:1000}. Activation of this defect according
to $[$Ti$^{3+}$--$\mathrm{V_{O}}$--Ti$^{3+}$$]$ $\rightarrow$
$[$Ti$^{3+}$--$\mathrm{V_{O}}$$]$ + Ti$^{3+}$ may indeed cause a
$\chi_{\mathrm{M}}$ anomaly similar to that observed in
Fig.\ref{fig1}. However, donor doping always decreases the
concentration of $\mathrm{V_{O}}$ \cite{lewis:1986}. Hence one would
expect a decrease in the $\chi_{\mathrm{M}}$ anomaly with increasing
doping, which is opposite to what is seen in our results. Finally,
we rule out the possibility that the $\chi_{\mathrm{M}}$ anomaly is
associated with a change in the ionization state of Nb$^{5+}$
according to Nb$^{4+}$ + Ti$^{4+}$ $\rightarrow$ Nb$^{5+}$ +
Ti$^{3+}$ since Fig.\ref{fig1b} shows similar magnetic anomalies
occurring when electrons are introduced into \bt\ by donors having a
stable oxidation state (Y$^{3+}$ and La$^{3+}$).

Recently, magnetic susceptibility measurements have been made on
several 3d$^1$ titanates including MgTi$_2$O$_4$, NaTiSi$_2$O$_6$
and LiTiSi$_2$O$_6$ \cite{isobe2:2002}. It was found that magnetic
anomalies in all the above compounds were associated with the
formation of spin-singlet Ti$^{3+}$--Ti$^{3+}$ pairs. In contrast to
the above compounds, the formation of BPs in n-type \bt\ is not
driven by a change in the crystal symmetry, as revealed in the
samples with \emph{x} $\ge$ 0.1. It also appears that pairing of the
electrons occurs even for very diluted 3d$^{0+\delta}$ ($\delta$
$\ll$ 1) electronic concentrations as in the case of \emph{x} =
0.002.
\begin{figure}[htb]
 \begin{center}
\scalebox{0.8}{\includegraphics*{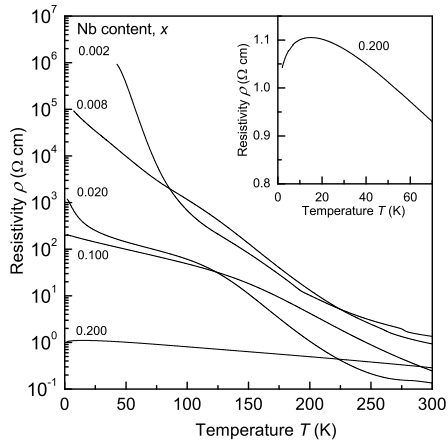}}
  \caption{Electrical resistivity of the BaTi$_{1-x}$Nb$_x$O$_3$ ceramics with \emph{x} = 0.002, 0.008, 0.020,
  0.100 and 0.200. The inset highlights the crossover from band to hopping conduction in the \emph{x} = 0.200 sample.}
  \label{fig2}
 \end{center}
\end{figure}

In view of the heated debate regarding the possibility of the
existence of \emph{mobile} small BPs \cite{chakraverty:1998}, we
have also examined the electronic behavior of n-type \bt. According
to the resistivity data shown in Fig.\ref{fig2}, the ground state of
n-type \bt\ is insulating throughout the 0 $<$ \emph{x} $<$ 0.2
doping range \cite{comment1:2005} and therefore comprises immobile
small BPs. The total conductivity can be expressed as a sum of
polaronic, $\mathrm{\sigma_{P}}$, and bipolaronic,
$\mathrm{\sigma_{BP}}$, conductivities
\begin{equation}
\sigma = \mathrm{\sigma_{P}} + \mathrm{\sigma_{BP}} =
ec_{\mathrm{P}} \mu_{\mathrm{P}}
+2e\frac{x-c_{\mathrm{P}}}{2}\mu_{\mathrm{BP}}, \label{eq5}
\end {equation}
where $\mu$$_{\mathrm{P}}$ and $\mu$$_{\mathrm{BP}}$ are the drift
mobilities of SPs and BPs, respectively. We assume, in accordance
with Ref. \cite{emin:1996}, that for a low doping level the main
contribution to $\sigma$ comes from the SPs that are formed by
thermal dissociation of the BPs and that $\mathrm{\sigma_{BP}}$ is
negligible due to the large effective mass and very low mobility of
BPs. Then, in the case of non-adiabatic hopping of SPs
\cite{bursian:1971},
\begin{equation}
\sigma \approx ec_{\mathrm{P}} \mu_{\mathrm{P}} \propto
\frac{1}{T^{1.5}}\mathrm{exp} \left[-\frac{\varepsilon_{\mathrm{b}}
+ E_{\mathrm{th}}}{k_{\mathrm{B}}T} \right]. \label{eq6}
\end {equation}

\begin{figure}[htb]
 \begin{center}
\scalebox{1}{\includegraphics{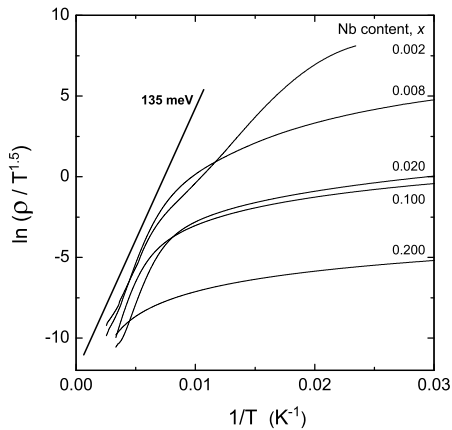}}
  \caption{Arrhenius plot of the resistivity of the BaTi$_{1-x}$Nb$_x$O$_3$ ceramics assuming non-adiabatic hopping of the localized carriers.}
  \label{fig3}
 \end{center}
\end{figure}

Taking \emph{E}$_{\mathrm{th}}$ = 23 meV from Ref. \cite{ihrig:1976}
and $\varepsilon$$_{\mathrm{b}}$ from Table \ref{table1}, the total
activation energy of conductivity, $\varepsilon$$_{\mathrm{b}}$ +
\emph{E}$_{\mathrm{th}}$, should be around 135 $\pm$ 10 meV at high
temperatures, with a slight deviation for the \emph{x} = 0.2 sample.
Indeed the resistivity data confirm that all but the \emph{x} = 0.2
sample show activated conductivity above \emph{T} $\approx$ 155 K
with an activation energy of ca. 135 meV as shown in Fig.\ref{fig3}.
The strong downturn from Arrhenius behavior for \emph{T} $<$ 155 K
is due to an enhancement of the mobility by quantum lattice
fluctuations \cite{fratini:2003}.
As the dopant concentration increases, the $\mathrm{\sigma_{BP}}$
contribution can no longer be neglected. In fact, the sample with
the highest doping level (\emph{x} = 0.2) shows a strong
conductivity enhancement at low temperature, with a maximum in
$\rho$ at \emph{T} = 15 K as shown in the inset of Fig.\ref{fig2}.
According to the SP theory, the resistivity maximum is expected
\cite{fratini:2003} but rarely observed in real polaronic systems
\cite{schein:1978}. In our case, it is associated with a crossover
from the coherent transport of BPs within the bipolaron band below
15 K to the incoherent (i.e., hopping) transport of BPs above 15 K.
We should stress here that this phenomenon is not unique to Nb-doped
samples and has also been observed in La-doped \bt.

Finally, let us compare the thermal dissociation energy of BPs,
$\varepsilon$$_{\mathrm{b}}$, obtained in this work with literature
data on the infra-red absorption band of n-type \bt\ centered at
\emph{E}$_{\mathrm{op}}$ = 500 -- 600 meV
\cite{berglund:1967a,bursian:1971}. Optical absorption of BPs
involves the splitting of the two electrons and transfer of one of
them to the next neighboring site. According to Emin
\cite{emin:1993}, this process would require an energy of
\emph{E}$_{\mathrm{op}}$ $\approx$ 4$\varepsilon$$_{\mathrm{b}}$.
Taking $\varepsilon$$_{\mathrm{b}}$ data for weakly doped samples
(\emph{x} $<$ 0.02) from Table \ref{table1}, the optical absorption
due to the BPs would produce a band centered at 420 -- 490 meV,
which is in reasonable agreement with the literature data. A
BP-associated optical absorption is expected to be strongly
temperature dependent since the concentration of the BPs decreases
with temperature. An alternative source of the mid-infrared
absorption band in n-type \bt\ could be the d$_{xy}$ $\rightarrow$
d$_{xz}$, d$_{yz}$ orbital (intra-band) excitations. Formation of
the SP removes the three-fold t$_{2g}$ orbital degeneracy in \bt\
due to the \emph{T}$_2$$\times$\emph{e} Jahn-Teller effect.
According to ESR data, the energy associated with the orbital
excitations within the t$_{2g}$ band in n-type \bt\ is around 400 --
536 meV \cite{scharfschwerdt:1996a}. In this case, however, the
intensity and the energy maximum of the optical absorption should be
temperature independent. Since there are no data on the temperature
dependence of the mid-infrared band of n-type \bt, the certain
origin of this band remains unresolved.

In conclusion, as revealed by magnetic susceptibility data,
``itinerant'' electrons in BaTi$_{1-x}$Nb$_x$O$_3$ with 0 $<$
\emph{x} $<$ 0.2 form localized pairs with a spin singlet ground
state. The binding energy of the BPs ranges from 105 meV to 122 meV
depending on the dopant concentration. Electrical resistivity
measurements show that, throughout this doping range, the BPs are
\emph{immobile}. A crossover from band to hopping transport is
manifested by a resistivity maximum at \emph{T} = 15 K in the most
heavily doped (\emph{x} = 0.2) sample investigated. This indicates
that at rather high doping concentration, small BPs in \bt\ become
\emph{mobile}. In a more general perspective, we speculate that \bt\
and \st\ play host to the two extreme cases of the bipolaron ground
state: the small BP in \bt\ evolves into a large BP in \st. Hence,
\bt\ -- \st\ may become a model system for studies of the physics of
small and large BPs as it became a model system for studies of soft
mode behavior 50 years ago.
\begin{acknowledgments}

This study was performed using Special Coordination Funds for
Promoting Science and Technology from the Ministry of Education,
Culture, Sports, Science and Technology of the Japanese Government.
The authors are grateful to I. Solovyev, A. Mishchenko and D.
Khomskii for helpful comments.

\end{acknowledgments}


\begin{thebibliography}{99}
\bibitem{comment1:2005} In accordance
with the resistivity data, no electronic contribution to the
specific heat has been detected at low temperatures, indicating the
absence of itinerant carriers.
\end{thebibliography}
\end{document}